\documentclass[pdflatex,sn-basic]{sn-jnl}

\usepackage{graphicx}
\usepackage{multirow}
\usepackage{amsmath,amssymb,amsfonts}
\usepackage{amsthm}
\usepackage{mathrsfs}
\usepackage[title]{appendix}
\usepackage{xcolor}
\usepackage{textcomp}
\usepackage{manyfoot}
\usepackage{booktabs}
\usepackage{float}
\usepackage{algorithm}
\usepackage{algorithmicx}
\usepackage{algpseudocode}
\usepackage{listings}

\usepackage{bm}
\usepackage{mathtools}
\usepackage{physics}
\usepackage{nicefrac}
\usepackage{enumitem}
\usepackage{textgreek}
\usepackage{cleveref}

\makeatletter
\newcommand{\printfnsymbol}[1]{\textsuperscript{\@fnsymbol{#1}}}
\makeatother

\theoremstyle{thmstyleone}

\theoremstyle{thmstyletwo}

\theoremstyle{thmstylethree}

\begin{document}

\title[Creep Topology Optimization]{Thermo-Structural Topology Optimization Considering Nonlinear Creep}

\author[1]{\fnm{Stefan} \sur{Knapik}}\email{stefan.knapik@northwestern.edu}
\equalcont{These authors contributed equally to this work.}

\author[1]{\fnm{Aaditya} \sur{Chandrasekhar}}\email{aadityacs@northwestern.edu}
\equalcont{These authors contributed equally to this work.}

\author[1]{\fnm{Deepak} \sur{Sharma}}\email{deepak.sharma@northwestern.edu}

\author[1]{\fnm{Jian} \sur{Cao}}\email{jcao@northwestern.edu}

\author[2]{\fnm{Changjie} \sur{Sun}}\email{changjie.sun@geaerospace.com}

\author*[1]{\fnm{Wei} \sur{Chen}}\email{weichen@northwestern.edu}

\affil[1]{\orgdiv{Department of Mechanical Engineering}, \orgname{Northwestern University},
  \orgaddress{\city{Evanston}, \state{IL}, \postcode{60208}, \country{USA}}}

\affil[2]{\orgname{GE Aerospace Research},
  \orgaddress{\city{Niskayuna}, \state{NY}, \postcode{12309}, \country{USA}}}

\abstract{Creep is a primary life-limiting mechanism for metallic components operating at high temperature, producing permanent deformation under sustained loads even when stresses remain below yield. The design of structures to minimize this deformation is critical to extending the service life of components. Incorporating creep into topology optimization (TO) remains open because the response is nonlinear, history-dependent, and thermomechanically coupled, and prior work often relies on linear viscoelastic models, which do not capture the behavior of metals at high temperatures. To bridge this gap, we introduce a differentiable thermo-structural TO framework. The approach considers creep deformation using the Norton model and leverages JAX's automatic differentiation to perform adjoint sensitivity analysis, enabling efficient gradient-based optimization. The transient material response is solved via a backward Euler scheme over a prescribed service life. Our objective is to minimize creep deformation subject to a volume constraint. We first demonstrate the framework on canonical two-dimensional benchmarks, showing that the proposed formulation significantly reduces permanent deformation compared to designs optimized solely for elastic stiffness. We then pose, as a challenge problem, the compositional design of a three-dimensional graded material turbine blade in which the local mixture of two candidate alloys is optimized. This challenge problem exercises the full capability of the framework, including transient nonlinear creep, coupled thermal loading, three-dimensional geometry, and gradient-based multi-material design, highlighting the need for creep-aware design in high-temperature applications.}


\maketitle

\section{Introduction}
\label{sec:intro}

Components such as turbine blades, combustor liners, and heat exchangers often operate under large structural loads at elevated temperatures \citep{yuan2023reviewCreepAircraft}. These conditions often lead to material creep; a critical failure and life-limiting factor. Creep is a gradual, time-dependent, permanent deformation of a material under sustained stress, which occurs over extended periods even at stress levels below the material's yield stress \citep{bailey1930creep}. Designing structures to effectively minimize creep deformation is therefore essential for ensuring structural integrity and extending the service life.

The design of creep-resistant structures, however, presents significant challenges. Experimental studies are often limited, as creep occurs over long durations under conditions difficult to replicate in a lab \citep{wu2025creep}. Furthermore, creep is a complex phenomenon characterized by transient, nonlinear, and multi-physics mechanisms \citep{naumenko2007modelingCreep}. These difficulties motivate simulation-driven design, where structures are optimized computationally rather than through experimental iteration. Among such approaches, topology optimization (TO) has emerged as a particularly powerful one \citep{deaton2014survey}. TO methods optimize the material distribution subject to specified objectives and constraints. This allows them to navigate complex design landscapes that challenge conventional, intuition-based approaches.

While TO is well-suited for complex structural design, its application to creep has been limited. Accurately capturing creep requires nonlinear, time-dependent constitutive models \citep{chen2026MLCreep}, which present considerable design challenges. Prior work on creep-aware TO has predominantly used linear viscoelastic models \citep{james2015TOViscoCreep,ogawa2022minimizingCreep}. Such models neglect the critical dependence of creep on temperature and are primarily applicable to polymers. They fall short in representing the behavior found in high-temperature applications where creep is inherently coupled with thermal effects.

To address these limitations, we propose a thermo-structural TO framework that integrates transient, temperature-dependent, nonlinear creep behavior using the Norton constitutive model \citep{norton1929creep}. The Norton model captures the secondary (steady-state) creep regime that governs long-term deformation in metallic components at elevated temperatures, providing a physically faithful yet tractable description of the temperature-coupled, time-dependent response that the simplified viscoelastic models used in prior work cannot represent. The framework is implemented in an end-to-end differentiable environment using JAX \citep{jax2018github}, which enables automatic computation of design sensitivities without manual adjoint derivations. An implicit backward Euler time-integration scheme keeps memory requirements low by permitting stable coarse time steps. By directly incorporating creep into the optimization objective, subject to a volume constraint, our framework systematically minimizes long-term permanent deformation in high-temperature applications while keeping the structure lightweight. We demonstrate the framework on several thermo-structural design optimization problems and show that the thermal coupling qualitatively changes the optimized topology relative to isothermal creep designs.

The remainder of this paper is organized as follows. \Cref{sec:relatedWork} details some of the prior literature in this domain. We follow this by a description of the governing thermo-structural equations and the Norton creep model in \Cref{sec:method_govEq_thermoElastic} and \Cref{sec:method_govEq_creep} respectively. \Cref{sec:method_optimization} describes the TO problem formulation and the sensitivity analysis scheme. We then present validation, Pareto trade-off, thermal-coupling, and load magnitude parametric studies in \Cref{sec:expts}, followed by a three-dimensional compositionally-graded turbine blade design in \Cref{sec:expts_3D}. Finally, we present some concluding remarks in \Cref{sec:discussion}.

\section{Related Work}
\label{sec:relatedWork}

Given our central contribution of developing a differentiable, thermo-structural optimization framework for creep deformation, we structure this review into key areas: the creep phenomenon, constitutive creep models, and recent advances in creep-aware structural optimization.

Over the years, topology optimization (TO) has evolved from its origins in stiffness maximization \citep{bendsoe1988generating} and stress-constrained design \citep{holmberg2013stress} to encompass thermal \citep{gao2008topology}, fluid \citep{borrvall2003topology}, and multiphysics \citep{alexandersen2014topology, yu2020three} problems, as well as increasingly complex material models. Early work on TO with inelastic materials focused primarily on plasticity and damage. \citet{mayer1996application} introduced an elastoplastic TO for crashworthiness, while \citet{swan1997voigt} presented a pioneering formulation that included viscoelasticity alongside plasticity, demonstrating the feasibility of optimizing structures with time-dependent behavior. \citet{maute1998adaptive} subsequently developed a formulation based on the von-Mises yield criterion to optimize structures for ductility.

Building on these advances, researchers began exploring nonlinear damage models within TO. \citet{desmorat2008topology} used a damage-based approach to model plasticity due to cyclical loading. Such work required path-dependent adjoint sensitivity analyses to handle the history-dependent material response. Subsequent advancements have significantly matured the integration of TO with plasticity, with numerous studies focusing on optimizing structural geometries with a single material \citep{maute1998adaptive, wallin2016topology, alberdi2017topology, kim2020microstructure, ivarsson2021plastic}. Amidst these advances in plasticity and damage, TO for creep remains an underexplored but critical area for high-temperature applications, a gap addressed in this paper.

Creep is the time-dependent, permanent deformation of materials under a sustained load, and becomes significant at elevated temperature or under prolonged loading. For metals, especially high-melting-point alloys and superalloys, creep deformation is typically negligible at ambient conditions and becomes a design concern when the temperature exceeds roughly 0.4-0.5 of the melting temperature \citep{frost1982deformation}. In contrast, many polymers and plastics can creep at room temperature because their glass-transition or softening temperatures are comparatively low, so components such as plastic fixtures may slowly deform under sustained load \citep{ferry1980viscoelastic, ward1993introduction}. This temperature dependence is central to engineering design: in high-temperature metal applications such as gas-turbine hot-section components and heat exchangers, service temperatures commonly reach 500-1200 $^{\circ}C$, and creep-driven elongation or distortion must be controlled to avoid closing clearances or loss of dimensional accuracy \citep{pollock2006nickel}, while polymer components used at or above their softening temperature require creep-aware design even at ambient conditions \citep{ward1993introduction}. Recognizing when a structure operates at a substantial fraction of its melting point (metals) or softening point (polymers) guides whether creep must be explicitly considered in analysis and optimization \citep{frost1982deformation}. Addressing this temperature-dependent design consideration for metals is a central focus of the present work.

Modeling creep behavior is essential for predicting the long-term performance of structures. Several empirical and mechanics-based constitutive models have been developed to characterize creep strain accumulation. One widely used formulation is Norton's power-law creep model, which expresses the steady-state creep strain rate as a power function of stress \citep{norton1929creep}. The Norton model effectively captures the secondary (steady-state) creep regime \citep{bailey1930creep}, which is of major interest in structural component design as the primary creep has a minor contribution to the overall deformation (e.g., 0.05\% to 0.2\% of primary creep strain for Ni-based superalloys). 

Building on this, the Norton–Bailey creep law introduces an explicit time dependence to also capture primary creep behavior. In this model, creep strain is proportional to both a power of time and a power of stress. Differentiating this relation yields a creep strain-rate formula that includes a time term, often called a time-hardening formulation. Another class of creep models incorporates continuum damage mechanics (CDM) to represent material degradation. A prominent example is the Kachanov-Rabotnov damage model, one of the earliest CDM-based creep formulations \citep{kachanov1999rupture, rabotnov1970creep, lemaitre1994mechanics}. This model introduces a damage variable into the constitutive equations, which evolves as creep strain accumulates. The model uses a set of coupled equations to describe how creep strain and damage progress simultaneously. As damage grows toward its fully-degraded limit, the material's effective load-bearing capacity is reduced, leading to an acceleration of creep. This captures the tertiary creep stage and eventual creep rupture, making the model valuable for predicting the time to failure in components such as pressure vessels. We choose a Norton-type creep model in this work due to its effectiveness in capturing the dominant secondary creep stage, which is often the primary concern for long-term deformation analysis.

While TO has traditionally focused on static performance, recent work has begun incorporating time-dependent deformation into the optimization, enabling structures to remain performant over their intended lifespans. One of the pioneering contributions in this area was by \citet{james2015TOViscoCreep}, who developed a TO framework for structures undergoing viscoelastic creep, using a time-dependent adjoint method to compute sensitivity derivatives. Their approach integrated a linear viscoelastic finite element model into the optimization loop, allowing the design to be optimized for a specified service duration with the goal of minimizing mass while satisfying long-term deflection limits.

Building on such ideas, several recent studies have advanced the inclusion of creep in TO. \citet{takalloozadeh2019topology} examined TO under high-temperature creep by considering stress relaxation. They implemented a time-dependent adjoint sensitivity analysis to include creep's stress-redistribution effect in a compliance-minimization framework. Following that, \citet{yoon2023stress} extended creep-aware TO to a stress-based formulation. They tackled the problem of minimizing the maximum von Mises stress while including creep effects, developing a time-dependent adjoint method tailored to the maximum stress criterion. Most recently, \citet{ogawa2022minimizingCreep} introduced a TO approach explicitly aimed at minimizing creep deformation. Their formulation directly targeted the time-dependent displacement due to creep as the objective. They adopted a generalized Maxwell model to represent creep behavior and derived the sensitivity of the creep deformation with respect to the material distribution, which is non-trivial because the objective depends on the entire time history of the response.

Despite these advances, existing creep-aware TO frameworks, including \citet{james2015TOViscoCreep,ogawa2022minimizingCreep}, rely on linear viscoelastic or Maxwell-type models that do not capture the temperature dependence of creep. They are therefore ill-suited for metallic components operating at elevated temperatures, where the Arrhenius sensitivity of the creep rate to temperature is a first-order effect. No prior work has formulated a TO framework that couples a steady-state thermal solve with a transient nonlinear creep model (e.g., Norton's power law) and optimizes material distribution to minimize creep-induced deformation under combined thermo-mechanical loading. This gap motivates the present work.

\section{Proposed Method}
\label{sec:method}
\subsection{Problem Overview}
\label{sec:method_problemOverview}

\begin{figure}[htbp]
    \centering
    \includegraphics[width=0.5\linewidth]{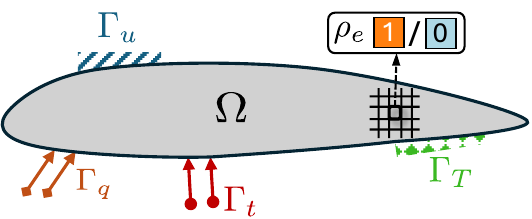}
    \caption{Design domain $\Omega$ under thermo-mechanical loading: heat flux on $\Gamma_q$, temperature on $\Gamma_T$, traction on $\Gamma_t$, and displacement on $\Gamma_u$. Material distribution is parameterized by elemental pseudodensities $\rho_e \in (0, 1]$ for topology optimization.}
    \label{fig:method_domain}
\end{figure}

As discussed in \Cref{sec:relatedWork}, existing creep-aware TO frameworks rely on linear viscoelastic or Maxwell-type constitutive models that do not accurately capture the nonlinear creep behavior of high-temperature metals, whose creep strain rate depends nonlinearly on both stress and temperature. To address this gap, we propose a gradient-based topology optimization (TO) framework that minimizes thermo-structural creep deformation under an imposed volume constraint. We begin the formulation assuming a prescribed design domain $(\Omega)$ and boundary conditions visualized in \Cref{fig:method_domain}. Furthermore, we assume prescribed materials with the appropriate thermal, elastic, and (Norton) creep properties. Our objective is to compute an optimal design by determining the ideal material distribution at each spatial location within $\Omega$. Notably, we require the distribution to minimize the deformation due to creep while being lightweight.

We begin by discussing the design representation in \Cref{sec:method_designRepresentation}. We then detail the governing physics and corresponding finite element formulations in \Cref{sec:method_govEq}. The optimization problem including the design variables, objective and constraints are presented in \Cref{sec:method_optimization}, followed by an algorithmic overview of the complete framework in \Cref{sec:method_overview}. The full pseudocode for the design loop, transient analysis, and constitutive update is provided in \Cref{sec:appendix_algorithms}.

\subsection{Design Representation}
\label{sec:method_designRepresentation}

We employ standard element-wise pseudodensities \citep{sigmund200199} to represent the design. In particular, we begin by discretizing our design domain $\Omega$ into finite elements, as illustrated in \Cref{fig:method_domain}. We then assign to each element $e$ a pseudodensity $\rho_e \in [\rho_{\min}, 1]$, where a value of one indicates the element is filled with solid material and a value of $\rho_{\min}$ indicates a near-void. In our experiments, we set $\rho_{\min} = 0.001$ to prevent a singular stiffness matrix while maintaining a negligible contribution from void elements. Although the pseudodensities may assume any continuous value in $[\rho_{\min}, 1]$, we require a near-binary design at convergence. Additionally, we impose density filtering to prevent checkerboard patterns from emerging during optimization \citep{diaz1995checkerboard}. In particular, we use a radial filter:

\begin{equation}
\tilde{\rho}_e =
\frac{
\sum\limits_{i} \omega_{i,e}\,\rho_i
}{
\sum\limits_{i} \omega_{i,e}
} \quad , \quad \omega_{i,e} = \max\!\left(r_f - \|\bm{x}_i - \bm{x}_e\|_2,\; 0\right)
\end{equation}

where $r_f$ is an imposed filter radius. We set $r_f$ equal to 1.5 times the mean element size. The filtered density $\tilde{\rho}_e \in [\rho_{\min}, 1]$ serves as the physical pseudodensity and is passed directly to the finite element analysis to evaluate material properties.

\subsection{Governing Equations and Discretization}
\label{sec:method_govEq}

\subsubsection{Thermo-Elastic Analysis}
\label{sec:method_govEq_thermoElastic}

This study focuses on the TO of structures undergoing creep under thermo-mechanical loads. We begin by considering the thermal problem. For simplicity, we assume a linear material with static thermal loads and response. The temperature field $T$ is governed by the steady-state heat conduction PDE:

\begin{subequations}
\label{eq:thermal_PDE}
\begin{align}
\bm{R}_T &\coloneqq -q_{j,j} + s = 0 \quad \text{in } \Omega  \\
q_j &= -\kappa T_{,j} \quad \text{in } \Omega  \\
T &= \bar{T} \quad \text{on } \Gamma_T \\
q_j n_j &= \bar{q} \quad \text{on } \Gamma_q
\end{align}
\end{subequations}

where $q_j$ is the heat flux vector, $s$ is a body heat source, and $\kappa$ is the thermal conductivity. $\bar{T}$ is the prescribed temperature on $\Gamma_T$, and $\bar{q}$ is the prescribed heat flux on $\Gamma_q$, as illustrated in \Cref{fig:method_domain}.

Furthermore, we consider a weakly coupled thermo-elastic system where temperature changes induce thermal strains. We assume the system operates within the small-strain regime and follows a linear Hookean constitutive law. The structural response is governed by:

\begin{subequations}
\label{eq:struct_PDE}
\begin{align}
\bm{R}_u &\coloneqq \sigma_{ij,j} + b_i = 0 \quad \text{in } \Omega \\
\sigma_{ij} &= \mathbb{E}_{ijkl}\varepsilon_{kl}^{el}  \\
u_i &= \bar{u}_i \quad \text{on } \Gamma_u \\
\sigma_{ij}n_j &= \bar{t}_j \quad \text{on } \Gamma_t
\end{align}
\end{subequations}

Here, $\sigma_{ij}$ is the Cauchy stress tensor, $b_i$ is the body force vector, and $\mathbb{E}_{ijkl}$ is the elasticity tensor. $\bar{u}_i$ and $\bar{t}_j$ are the prescribed displacement and traction on boundaries $\Gamma_u$ and $\Gamma_t$, respectively, as illustrated in \Cref{fig:method_domain}. For simple thermo-elastic deformation, the total strain is additively decomposed into elastic and thermal components, giving the elastic strain $\varepsilon_{ij}^{el} = \varepsilon_{ij}^{\Sigma} - \varepsilon_{ij}^{th}$, where $\varepsilon_{ij}^{\Sigma} = \frac{1}{2}(u_{i,j} + u_{j,i})$ is the total strain and $\varepsilon_{ij}^{th} = \alpha(T - T_0)\delta_{ij}$ is the thermal strain. Here, $\alpha$ is the coefficient of thermal expansion and $T_0$ is the reference temperature at which thermal strains vanish. However, for creep-induced deformation, an additional inelastic component must be introduced into the strain decomposition, as detailed next.

\subsubsection{Creep Model}
\label{sec:method_govEq_creep}

Having established the thermo-elastic governing equations, we now turn to the primary focus of this work: designing structures that minimize deformation due to creep. Creep is characterized by gradual, permanent deformation that accumulates under sustained stress, notably even when the stresses remain below the material's yield stress. As discussed in \Cref{sec:relatedWork}, several constitutive models have been proposed to characterize creep, including the Norton-Bailey time-hardening law \citep{bailey1930creep} to capture primary and secondary creep, and the Kachanov-Rabotnov damage model for tertiary creep \citep{kachanov1999rupture}. Our focus is on designing metallic components operating in high-temperature environments. We therefore adopt the Norton power-law creep model \citep{norton1929creep}, which effectively characterizes the dominant secondary steady-state creep regime without strain hardening.

In high-temperature applications, the material's susceptibility to creep depends strongly on temperature. We capture this via an Arrhenius-type relationship for the creep coefficient $A(T)$:

\begin{equation}
    A(T) = A_0 \exp\!\left(\frac{-Q}{RT}\right)
    \label{eq:norton_creep_coeff}
\end{equation}

where $A_0$ is the pre-exponential factor, $Q$ is the activation energy, and $R$ is the universal gas constant. The parameters $A_0$ and $Q$ are material-specific constants. Consistent with this equation, higher temperatures accelerate creep, while high activation energies confer creep resistance at elevated temperatures.

In contrast to the thermoelastic decomposition of \Cref{sec:method_govEq_thermoElastic}, creep introduces an additional inelastic, time-dependent component. Retaining the small-strain assumption, the total strain $\varepsilon_{ij}^{\Sigma}(t)$ is additively decomposed into elastic, inelastic creep, and thermal components \citep{de2008computationalPlasticity}:

\begin{equation}
    \varepsilon_{ij}^{\Sigma}(t) = \varepsilon_{ij}^{el}(t) + \varepsilon_{ij}^{cr}(t) + \varepsilon_{ij}^{th}
    \label{eq:total_strain_creep}
\end{equation}

Since only the elastic component drives stress, the constitutive relation is:

\begin{equation}
    \sigma_{ij}(t) = \mathbb{E}_{ijkl} \left( \varepsilon_{kl}^{\Sigma}(t) - \varepsilon_{kl}^{cr}(t) - \varepsilon_{kl}^{th} \right)
    \label{eq:cauchy_stress_creep}
\end{equation}

The thermal strain $\varepsilon_{ij}^{th}$ is treated as quasi-static, consistent with the static thermal field assumption. The creep strain rate is governed by the Norton flow rule:

\begin{equation}
\dot{\varepsilon}_{ij}^{cr} = A(T)\,\sigma_{eq}^{n}\,N_{ij}, \qquad N_{ij} = \frac{3}{2}\frac{s_{ij}}{\sigma_{eq}}
\label{eq:norton_creep_strain_rate}
\end{equation}

where $s_{ij} = \sigma_{ij} - \frac{1}{3}\sigma_{kk}\delta_{ij}$ is the deviatoric stress, $\sigma_{eq} = \sqrt{\frac{3}{2}s_{ij}s_{ij}}$ is the von Mises equivalent stress, $N_{ij}$ is the flow direction tensor, and $n$ is the Norton stress exponent. The parameter $n$ is material-specific and governs the sensitivity of creep rate to stress magnitude.

To solve this transient problem, we discretize the time domain and employ an implicit backward Euler integration scheme. At time step $t_s$, given the previous creep strain $\varepsilon_{ij,s-1}^{cr}$ and current total strain $\varepsilon_{ij,s}$, we seek the creep strain increment $\Delta\varepsilon_{ij}^{cr}$ satisfying the implicit residual:

\begin{equation}
\mathcal{R}_{ij}\!\left(\Delta\varepsilon_{ij}^{cr}\right) \equiv \Delta\varepsilon_{ij}^{cr} - \Delta t_s\,\dot{\varepsilon}_{ij}^{cr}\!\left(\sigma_{ij,s}\right) = 0
\label{eq:creep_local_residual_tensor}
\end{equation}

where the stress $\sigma_{ij,s}$ is evaluated from the updated elastic strain:

\begin{equation}
\sigma_{ij,s} = \mathbb{E}_{ijkl}\!\left(\varepsilon_{kl,s} - \varepsilon_{kl,s-1}^{cr} - \Delta\varepsilon_{kl}^{cr} - \varepsilon_{kl}^{th}\right)
\label{eq:stress_creep_update}
\end{equation}

This tensor-valued nonlinear system couples $\Delta\varepsilon_{ij}^{cr}$ to the updated stress through the constitutive law. We resolve it at each Gauss point using Newton-Raphson iteration \citep{optimistix2024}. To avoid the laborious manual derivation of the Jacobian $\partial\mathcal{R}_{ij}/\partial\Delta\varepsilon_{kl}^{cr}$, and to make the framework readily extendable to other creep laws and constitutive models, we evaluate it using automatic differentiation. Once converged, the creep strain is updated as:

\begin{equation}
\varepsilon_{ij,s}^{cr} = \varepsilon_{ij,s-1}^{cr} + \Delta\varepsilon_{ij}^{cr}
\label{eq:creep_strain_update}
\end{equation}

A complete description of this local constitutive update is provided in \Cref{alg:local_update} (\Cref{sec:appendix_algorithms}).

\subsubsection{Finite Element Analysis}
\label{sec:method_govEq_FEA}

With the governing equations and constitutive model established, we now describe the finite element formulation. The thermal field is treated as steady-state and is evaluated prior to the time-dependent mechanical analysis. We solve the thermal problem first, and the resulting nodal temperatures serve as fixed inputs to the structural problem.

\paragraph{Thermal Problem.}
The domain is discretized into $N_e$ elements parameterized by filtered densities $\tilde{\rho}_e$ (\Cref{sec:method_designRepresentation}). This yields the discrete thermal system $\bm{K}_{th} \bm{T} = \bm{F}_{th}$, where $\bm{T}$ is the global vector of nodal temperatures. The global thermal stiffness matrix $\bm{K}_{th}$ is assembled from the elemental contributions:

\begin{equation}
  \bm{K}_{th}^e = \int\limits_{\Omega_e} \nabla \bm{N}^T \left( \kappa(\tilde{\rho}_e) \bm{I} \right) \nabla \bm{N}\, d\Omega_e
  \label{eq:thermal_stiffness_matrix}
\end{equation}

where $\bm{N}$ is the shape function matrix. Similarly, the thermal load vector $\bm{F}_{th}$ is obtained by assembling the elemental flux vectors $\bm{f}_{th}^e$, which account for the prescribed heat fluxes $\bar{q}$ on the boundary $\Gamma_q$ and internal heat sources $s$:

\begin{equation}
  \bm{f}_{th}^e = - \int\limits_{\Gamma_q^e} \bm{N}^T \bar{q}\, d\Gamma_q^e + \int\limits_{\Omega_e} \bm{N}^T s\, d\Omega_e
  \label{eq:thermal_load_vector}
\end{equation}

Upon solving this global thermal system, the resulting nodal temperatures $\bm{T}$ are interpolated via shape functions to the quadrature points of the structural problem, where the thermal strains $\varepsilon_{ij}^{th}$ and the Arrhenius creep coefficient $A(T)$ are evaluated at each quadrature point.

\paragraph{Structural Problem.}
Driven by the creep constitutive behavior, the ensuing structural response is inherently transient and nonlinear. Discretizing the time domain into $N_t$ steps, we employ a first-order implicit backward Euler integration scheme. For a given time step $t_s$, we seek the global displacement vector $\bm{u}_s$ that satisfies the discrete nonlinear equilibrium equation:

\begin{equation}
  \bm{R}(\bm{u}_s) = \bm{f}_{int}(\bm{u}_s, \bm{T}) - \bm{f}_{ext} = \bm{0}
  \label{eq:mechanical_residual}
\end{equation}

where $\bm{f}_{ext}$ is the external force vector. The global internal force vector $\bm{f}_{int}$ is assembled from the elemental contributions:

\begin{equation}
  \bm{f}_{int}^e = \int\limits_{\Omega_e} \bm{B}^T \bm{\sigma}_s(\bm{u}_e)\, d\Omega_e
  \label{eq:mechanical_internal_force}
\end{equation}

where $\bm{B}$ is the strain-displacement matrix and $\bm{\sigma}_s$ is determined by the local constitutive update (\Cref{alg:local_update}).

\paragraph{Material Penalization.}
To interpolate material properties as a function of density, we employ RAMP penalization \citep{stolpe_alternative_2001} with penalty parameter $q = 8$ for three material properties. For the thermal conductivity and elastic modulus, the standard RAMP form suppresses each property in void elements:

\begin{equation}
    \kappa(\tilde{\rho}_e) = \kappa_1 \cdot \frac{\tilde{\rho}_e}{1 + q(1 - \tilde{\rho}_e)}, \qquad
    E(\tilde{\rho}_e) = E_1 \cdot \frac{\tilde{\rho}_e}{1 + q(1 - \tilde{\rho}_e)}
    \label{eq:kappa_E_interpolation}
\end{equation}

$\kappa_1$ and $E_1$ are the thermal conductivity and elastic modulus of the solid material, respectively. This formulation ensures that void elements contribute negligibly to heat conduction and stiffness, while solid elements retain their full properties.

For the creep coefficient, we apply an inverted penalization:

\begin{equation}
    A_0(\tilde{\rho}_e) = \frac{A_{0,1}}{\left[\dfrac{\tilde{\rho}_e}{1 + q(1 - \tilde{\rho}_e)}\right]^{n}}
    \label{eq:A_interpolation}
\end{equation}

$A_{0,1}$ is the creep coefficient of the solid material. This formulation assigns a large creep coefficient to void elements, causing them to accumulate large creep strains under applied stress. This ensures that lesser densities exhibit less utility, which together with a volume constraint drive the design toward a binary density field. The exponent $n$ from the Norton model appears, and is chosen to exactly correct for the reduced stress elicted at a given strain in intermediate elements, such that creep strain rate is invariant to design density at given elastic strains. 


\paragraph{Global Equilibrium Solver.}
We resolve the global nonlinear equilibrium at each time step using a Newton-Raphson iterative solver, as outlined in \Cref{alg:global_nr}. At each iteration for a given time step $t_s$, the structural displacement increment $\Delta \bm{u}$ is obtained by solving the linearized system:

\begin{equation}
  \bm{K}_{T}(\bm{u}_s) \Delta \bm{u} = -\bm{R}(\bm{u}_s)
  \label{eq:mechanical_linearized_system}
\end{equation}

where $\bm{K}_{T} = \partial \bm{R}/\partial \bm{u}_s$ is the global tangent stiffness. Once global equilibrium converges, the nodal displacements and local creep strains are updated, and the simulation advances to $t_{s+1}$.

The solver is implemented with the JAX~\citep{jax2018github} library and builds on the work of JAX-FEM~\citep{xue_jax-fem_2023}, which leverages automatic differentiation for tangent stiffness matrix assembly. The entire simulation pipeline is end-to-end differentiable, facilitating the exact and efficient computation of sensitivities for gradient-based optimization.


\subsection{Optimization Formulation}
\label{sec:method_optimization}

Having established the design representation, governing equations, and finite element formulation, we outline the key components of the design optimization framework.

\paragraph{Design Variables.}
The material distribution is parameterized by elemental pseudodensities $\rho_e \in [\rho_{\min}, 1]$ with $\rho_{\min} = 0.001$. The raw design variables are subjected to a density filter (\Cref{sec:method_designRepresentation}) prior to finite element analysis to mitigate checkerboard patterns.

\paragraph{Objective.}
The primary goal of the proposed framework is to minimize the deformation induced by creep over the intended service life. Towards this, we define the objective as the work done by the applied loads acting through the accumulated creep displacement, analogous to a creep compliance measure. Noting that the structural simulation initializes with zero accumulated creep strain, the creep displacement can be isolated as the difference between the displacement at the final time $t_{N_t}$ and the initial purely elastic displacement: $\bm{u}^{cr} = \bm{u}_{N_t} - \bm{u}_0$. The integral of the external forces acting through this creep displacement on the traction boundary yields the objective:
\begin{equation}
    J = \int\limits_{\Gamma_t} \bm{u}^{cr} \cdot \bar{\bm{t}}\, d\Gamma_t 
  \label{eq:objective_creep_compliance}
\end{equation}

\paragraph{Volume Constraint.}
We impose a volume constraint to obtain lightweight designs. Let $V_*$ denote the maximum permissible volume fraction. The constraint is:


\begin{equation}
g_v = \frac{\int_{\Omega} \rho \, d\Omega}{V_* \int_{\Omega} d\Omega} - 1 \leq 0
\label{eq:volume_constraint_integral}
\end{equation}

\paragraph{Optimization Problem.}
Collecting the objective (\Cref{eq:objective_creep_compliance}), the thermal and structural PDEs (\Cref{eq:thermal_PDE,eq:struct_PDE}), and the volume constraint (\Cref{eq:volume_constraint_integral}), the complete optimization problem is:

\begin{subequations}
\label{eq:optimizationEquations}
\begin{align}
    & \underset{\bm{\rho}}{\text{minimize}}  & J  \label{eq:opt_objective} \\
    & \text{subject to} & \bm{R}_T = 0  \label{eq:opt_govnEqThermal} \\
    & & \bm{R}_u = 0  \label{eq:opt_govnEqStruct} \\
    & &  g_v \leq 0  \label{eq:opt_volCons} \\
    & &  \rho_{\min} \leq \bm{\rho} \leq 1  \label{eq:opt_partitionCons}
\end{align}
\end{subequations}

We employ the Method of Moving Asymptotes (MMA) \citep{svanberg1987MMA} as an optimizer for gradient-based design updates, using the Python implementation from \citep{deetman2025gcmma} with default parameters.

A critical step in gradient-based TO is computing sensitivities: the derivatives of the objective and constraints with respect to the design variables. This is particularly involved for transient, nonlinear multi-physics simulations. We construct an end-to-end differentiable pipeline using the automatic differentiation (AD) capabilities of JAX~\citep{jax2018github}, JAX-FEM~\citep{xue_jax-fem_2023}, and Optimistix~\citep{optimistix2024}, avoiding the laborious manual derivation of adjoint expressions. Because TO involves many design variables but only a few scalar outputs (the objective and constraints), we specifically employ reverse-mode AD, which propagates derivatives from each output back to all inputs in a single sweep; forward-mode AD, by contrast, would require a separate sweep per design variable and scale poorly as the design space grows.

One challenge inherent to our approach arises from iterative solvers (Newton-Raphson for global equilibrium and local stress updates). Naive AD would unroll derivatives through every solver iteration, which is computationally expensive. We instead apply the Implicit Function Theorem (IFT)~\citep{blondel2022ImplicitDifferentiation, Knapik2025}, automating the adjoint problem construction and avoiding Jacobian accumulation across each intermediate step of the Newton solvers (both global and local). This significantly reduces the memory requirement of reverse-mode AD while yielding machine-precise gradients.

\subsection{Algorithmic Overview}
\label{sec:method_overview}

\begin{figure}[p]
    \centering
    \includegraphics[height=0.95\textheight,keepaspectratio]{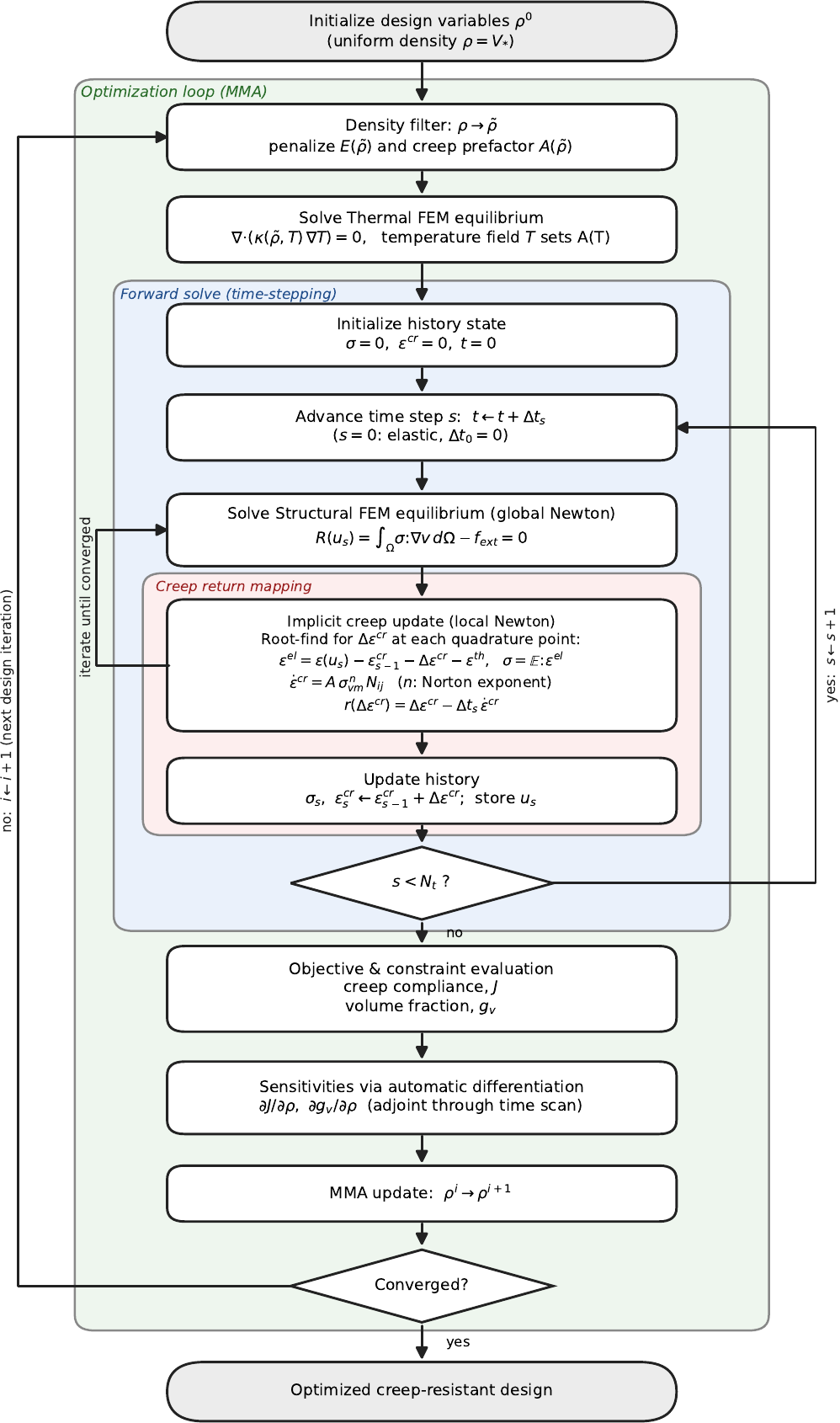}
    \caption{Algorithmic overview of the proposed creep topology optimization framework. Three nested loops are shown: the outer MMA design loop (green), the transient forward solve (blue), and the local creep return mapping at each Gauss point (red). Detailed pseudocode for each block is provided in \Cref{sec:appendix_algorithms}.}
    \label{fig:method_flowchart}
\end{figure}

The complete framework is summarized in \Cref{fig:method_flowchart}, which uses color coding to distinguish the principal nested loops. The outer MMA optimization loop (green) iterates over designs: at each iteration, the raw densities are filtered and penalized, the steady-state thermal problem is solved to obtain the nodal temperatures, and the transient creep response is evaluated. The intermediate time-stepping loop (blue) advances the structural state through $N_t$ implicit backward-Euler time steps. At each time increment, a global Newton-Raphson solver iterates to enforce equilibrium (\Cref{eq:mechanical_residual}), and within every one of its iterations the innermost local return-mapping loop (red) resolves the implicit creep residual (\Cref{eq:creep_local_residual_tensor}) at each Gauss point via Newton-Raphson with AD-computed Jacobians. Once the transient solve completes, the creep-compliance objective and volume constraint are evaluated, their sensitivities are obtained by reverse-mode AD combined with implicit differentiation through both Newton solvers, and MMA produces the next design. The procedure terminates when the design and objective change fall below a prescribed tolerance. Full pseudocode for each subroutine is provided in \Cref{sec:appendix_algorithms} (\Cref{alg:to_loop,alg:time_step,alg:global_nr,alg:local_update}).

\section{Numerical Experiments}
\label{sec:expts}

In this section, we conduct several experiments to illustrate the proposed framework. All experiments are conducted using JAX \citep{jax2018github} in Python, building upon the open-source JAX-FEM library \citep{xue_jax-fem_2023}. The default parameters used in the experiments are as follows:

\begin{enumerate}
  \item \textbf{Mesh and Elements :} We use a structured bilinear quad (Q4) mesh
  with $2 \times 2$ Gauss quadrature for 2D problems, and Hex8 elements with
  $2 \times 2 \times 2$ Gauss quadrature for a 3D problem. The default mesh for 2D examples is of a domain sized $200~\text{mm} \times 100~\text{mm}$ with a uniform resolution of $200 \times 100$ elements (element size $1~\text{mm} \times 1~\text{mm}$), and 2D problems are solved in plane stress with an assumed out-of-plane thickness of $1~\text{mm}$.
  
  \item \textbf{Optimizer :} The Method of Moving Asymptotes (MMA) is used for
  the design updates \citep{svanberg1987MMA}. The move limit is set to $10^{-1}$, and the default asymptotes
  initialization is used. The optimization loop is terminated either
  after a maximum of $250$ iterations or when the maximum change in design
  variables is $\|\bm{\rho}_{i} - \bm{\rho}_{i-1}\|_\infty \le 10^{-5}$.
  
  \item \textbf{Time Integration :} The transient creep analysis uses an implicit
  Backward Euler scheme. We discretize the service life into $N_t = 10$ time
  steps with a uniform time increment.
  
  

  \item \textbf{Material Properties :} Assuming isotropic elasticity, the solid material is assigned a
  Young's modulus of $E = 160.0$ GPa and a Poisson's ratio of $\nu = 0.3$, representative of a nickel-based superalloy (e.g., Inconel~625) at hot-section service temperature.
The Norton creep parameters are set to $A_0 = 10^{-21}~\mathrm{MPa}^{-n}\,\mathrm{s}^{-1}$ and $n = 3.5$. For purely structural problems, $Q = 0~\text{kJ/mol}$, while for thermo-structural problems, we set the activation energy to $Q = 10000~\text{kJ/mol}$ and the thermal conductivity to $k = 10~\text{W/mK}$. These Norton parameters are chosen as illustrative values to exercise the framework and are not calibrated to a specific alloy; alloy-specific parameters are introduced for the compositional study in \Cref{sec:expts_3D} (see \Cref{tab:blade_materials}). The single-material defaults listed above apply to the studies in \Cref{sec:expts_validation,sec:expts_pareto,sec:expts_thermal,sec:expts_loadmagnitude}.
\end{enumerate}

\subsection{Validation of the Creep-Aware Objective}
\label{sec:expts_validation}

 \begin{figure}[htb]
 	\begin{center}
		\includegraphics[width=1.0\linewidth]{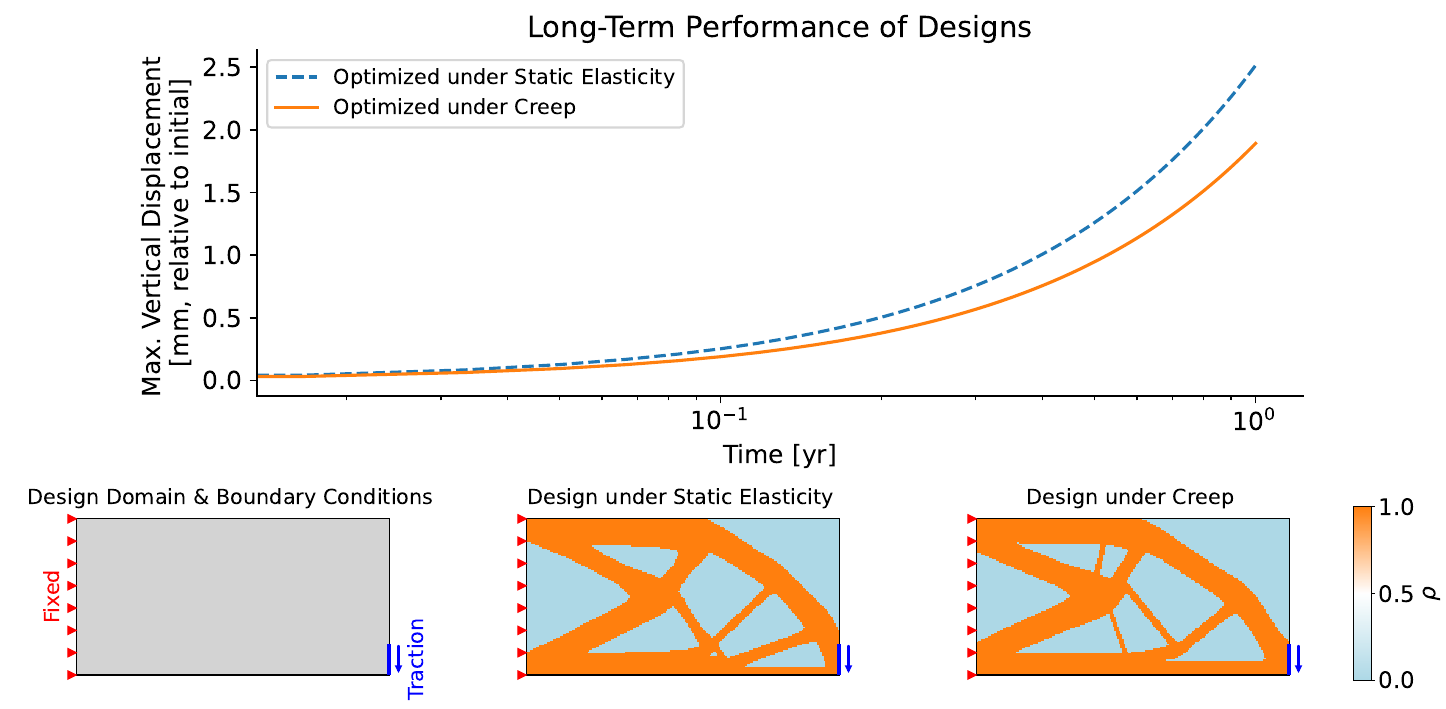}
 		\caption{Comparative analysis of creep-aware versus elastic design formulations, both optimized under an identical volume fraction limit of $V_* = 0.5$. The figure illustrates the problem setup (bottom left), the resulting topologies for both objectives (bottom right), and the evolution of tip displacement over the service life (top). The results highlight the divergence in long-term performance, with the creep-aware formulation yielding lower final displacements.}
 		\label{fig:result_validation}
	\end{center}
 \end{figure}

In this first experiment, we validate the central premise of the proposed framework: utilizing a creep aware objective leads to structures that outperform those designed under an elastic objective when subjected to long-term creep. To isolate the structural creep response, we consider a classic 2D cantilever beam subjected exclusively to sustained structural loads, neglecting thermal loads for this initial study. We use the default material properties and 2D mesh parameters outlined above, and apply a 100 MPa traction to the right edge of the beam with a service life of 1 year. We enforce a maximum allowed volume fraction of $V_* = 0.5$.

We perform two comparative studies. First, we consider the design subjected to the proposed creep aware objective defined in \Cref{eq:objective_creep_compliance}, which minimizes the external work done by the applied loads acting through the accumulated creep displacement. Next, we consider a baseline design that minimizes the initial elastic compliance at the first time step, neglecting creep accumulation entirely; $J_{el} = f_{ext}^T u_0$.

We observe (\Cref{fig:result_validation}) that while the optimized topologies appear visually similar at first glance, their long-term transient responses diverge considerably. The baseline design, optimized without creep awareness, initially exhibits a marginally lower elastic compliance but suffers from severe long-term deformation. In contrast, the creep-aware formulation mitigates these stress and time-dependent effects, resulting in a significantly reduced final tip displacement at the end of the service life. This motivates the consideration of creep during the design process, particularly for components expected to endure sustained loads.

Since the Norton creep rate scales nonlinearly with equivalent stress, peak-stress regions dominate the accumulated displacement, and the creep-aware optimizer reallocates material to suppress them, most visibly by thickening members near the fixed-boundary corners. The initial equivalent stress range narrows accordingly, from $0.04$ - $625$~MPa in the elastic design to $13.6$ - $474$~MPa in the creep-aware design. Under the shared volume budget, this reinforcement is offset by thinner members elsewhere, producing the fine truss features seen only in the creep-aware topology.

\subsection{Pareto Tradeoff}
\label{sec:expts_pareto}

 \begin{figure}[htb]
 	\begin{center}
		\includegraphics[width=0.8\linewidth]{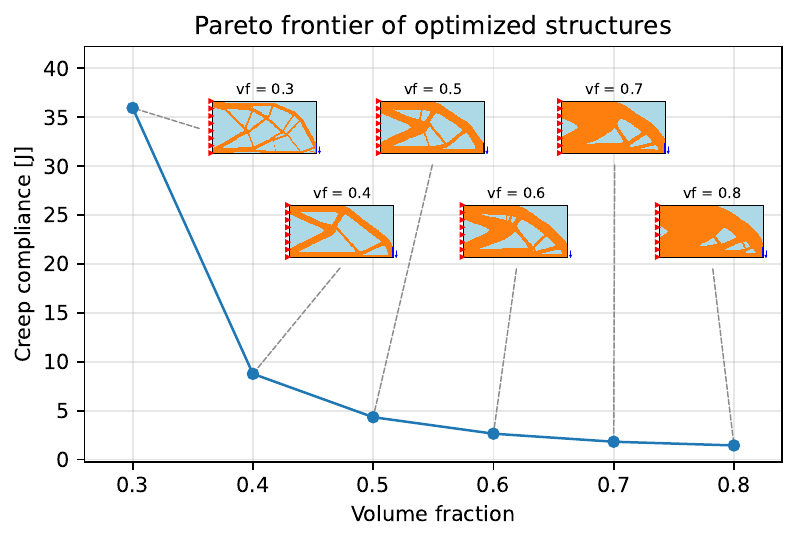}
 		\caption{Pareto tradeoff for the structural cantilever: creep compliance versus volume fraction $V_*$. As more material is permitted, the optimizer produces stiffer structures with reduced long-term deformation.}
 		\label{fig:result_pareto_structural}
	\end{center}
 \end{figure}

A fundamental consideration in structural design is the trade-off between performance and material usage. In this experiment, we study how creep compliance varies with the allowable volume fraction $V_*$, returning to the structural cantilever beam of \Cref{sec:expts_validation}.

We sweep the volume fraction constraint $V_*$ across a range of values while holding all other parameters constant. As anticipated, relaxing the volume constraint allows for stiffer structures; we observe a decrease in both the creep compliance and the final tip displacement as $V_*$ increases (\Cref{fig:result_pareto_structural}).

Creep compliance drops by roughly $75\%$ between $V_* = 0.3$ and $V_* = 0.4$, but flattens markedly beyond $V_* = 0.5$, indicating pronounced diminishing returns at higher volume fractions. This nonlinearity is a direct consequence of the Norton flow rule. Because the creep strain rate scales with stress raised to the power $n = 3.5$, even modest reductions in the peak sustained stress, enabled by additional material, yield disproportionately large reductions in the accumulated creep compliance. The accompanying topologies reflect this mechanism. At $V_* = 0.3$, the optimizer is forced to spread a scarce material budget across many thin, highly stressed members, and the resulting stress concentrations drive rapid Norton flow. As $V_*$ increases, the optimizer consolidates material into fewer, thicker load-bearing members that operate at lower sustained stress, and the topology transitions from a truss-like architecture toward a more compact, solid-body form. For this cantilever configuration, aggressive mass reduction incurs a disproportionate creep-life penalty, whereas modest material additions near the knee of the frontier ($V_* \approx 0.4$ - $0.5$) capture most of the achievable creep-compliance benefit.

\subsection{Effect of Thermal Loading on the Optimized Topology}
\label{sec:expts_thermal}

A central feature of the proposed framework is the explicit coupling between the thermal field and the creep response. In principle, temperature influences the structural problem through two mechanisms: (i) spatially varying thermal strains that alter the mechanical stress state, and (ii) spatial variation of the Arrhenius creep coefficient $A(T) = A_0\exp(-Q/RT)$, which concentrates creep damage in hotter regions. In this experiment we set the coefficient of thermal expansion, $\alpha = 0$, to isolate and demonstrate mechanism (ii) alone, keeping the stress state unchanged between cases so that all topological differences are attributable solely to the temperature-dependent creep coefficient. We conduct a controlled study on the 2D domain with left and right edges fully clamped and a downward traction of 100 MPa applied to the bottom edge over the span of 1 year. We use three temperature boundary condition configurations:

\begin{enumerate}
    \item \textbf{Low Temperature:} The entire domain is held at a uniform temperature $T = T_\text{cold} = 300~\text{K}$, equal to the cold-side reference. This serves as the baseline: the Arrhenius factor $\exp(-Q/RT)$ is spatially uniform and creep accumulates homogeneously.
    \item \textbf{High Temperature:} The entire domain is held at a uniform elevated temperature $T = T_\text{hot} = 1100~\text{K}$. The Arrhenius factor is amplified uniformly, increasing the overall creep rate, but introduces no spatial gradient in the creep coefficient.
    \item \textbf{Thermal gradient (thermo-structural):} A steady-state thermal gradient is imposed across the domain, cool on the left face ($T_\text{cold} = 300~\text{K}$) and hot on the right face ($T_\text{hot} = 1100~\text{K}$). Thermal conductivity governs the resulting temperature field. The spatially varying Arrhenius factor makes the hotter material more susceptiple to creep behavior.
\end{enumerate}

\begin{figure}[htb]
    \begin{center}
        \includegraphics[width=1.0\linewidth]{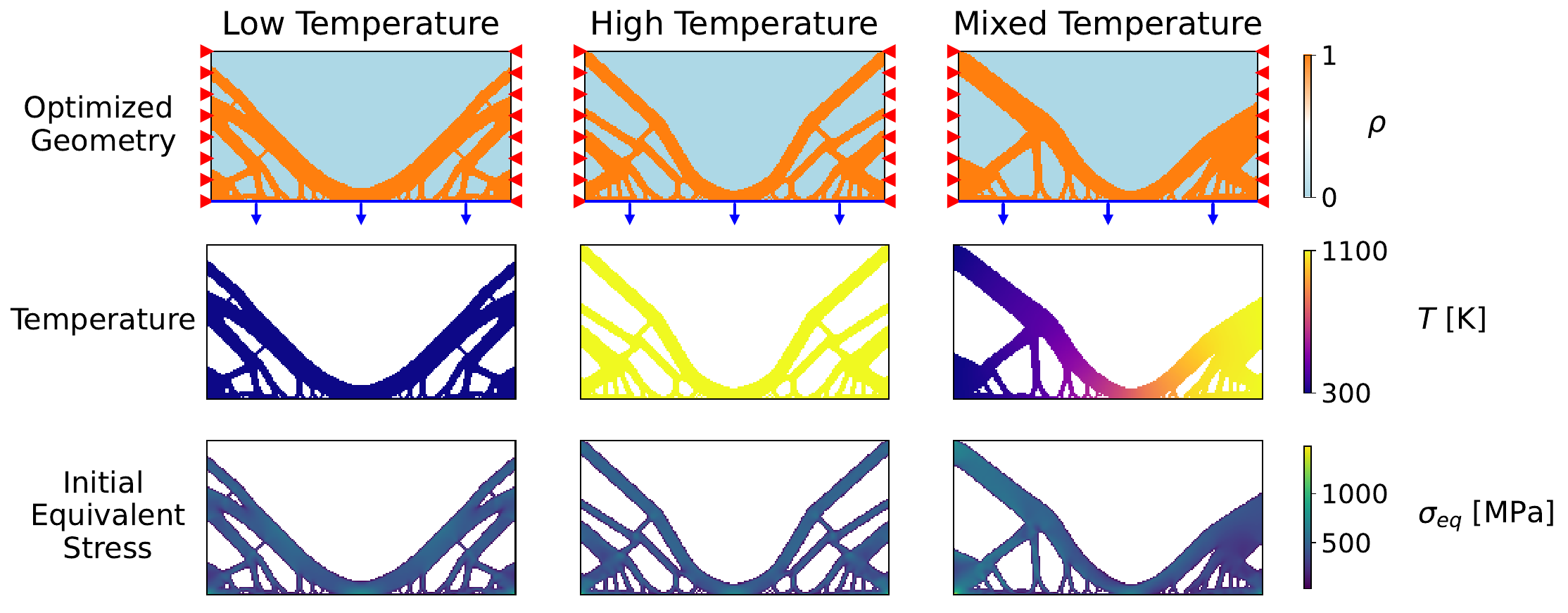}
        \caption{Effect of thermal loading on the optimized topology. \textit{Top}: Optimized material distributions for three loading configurations: low temperature, high temperature, and mixed temperature boundary conditions. \textit{Middle}: Spatial distribution of the temperature field. \textit{Bottom}: Spatial distribution of the initial ($t=0$) equivalent (von Mises) stress response, which drives the accumulation of creep strain.}
        \label{fig:result_thermoelastic}
    \end{center}
\end{figure}

All other parameters (traction loads, volume fraction, Norton exponent, service life) are held constant across the three cases, permitting a direct comparison of how thermal physics drives topology.

\Cref{fig:result_thermoelastic} shows the three optimized topologies alongside their initial von Mises stress distributions. Two observations stand out. First, the low- and high-temperature designs are both left-right symmetric and share the same overall truss-like architecture, though minor differences in member placement and thickness are visible. Because the Arrhenius factor is spatially uniform in both cases, it acts as a scalar amplifier of the creep rate and provides no directional cue to the optimizer; with $\alpha = 0$, the initial stress field is identical between the two, so the two problems differ only by a global scaling of the creep compliance. The residual differences in the recovered topologies reflect the non-convex optimization landscape and MMA sensitivity to objective magnitude rather than any physical redistribution of load. The practical takeaway is that a uniform temperature shift, in isolation, does not reshape the optimal material distribution; it only rescales the accumulated creep.

Second, the thermal-gradient case yields a visibly asymmetric topology: the optimizer thins the members on the cool left side and thickens the members on the hot right side. The stress panel in the bottom row makes the mechanism explicit, the thinner cool-side members carry higher sustained stress, while the thicker hot-side members operate at markedly lower stress. Because the Norton flow rate scales as $\dot{\varepsilon}_c \propto A(T)\,\sigma^n$ with $n = 3.5$, and $A(T)$ grows by several orders of magnitude between the cold and hot faces, the optimizer trades stress into the cool half (where $A(T)$ is small and elevated stress produces little creep) in order to trade stress out of the hot half (where $A(T)$ is large and any stress reduction is disproportionately valuable). The result is a creep-shielding of the thermally aggressive region, achieved by reinforcing the hot side rather than avoiding it.

The broader insight is that the Arrhenius temperature dependence of the creep coefficient is not a scalar correction to the mechanical problem but an active driver of the optimal material distribution whenever a spatial temperature gradient is present. Isothermal creep-aware designs, even those calibrated to the worst-case service temperature, will systematically misplace material in thermally inhomogeneous environments; capturing the coupled thermo-structural physics is essential for problems in which the temperature field varies over the design domain.

\subsection{Effect of Load Magnitude on the Optimized Topology}
\label{sec:expts_loadmagnitude}

The strong stress nonlinearity of the Norton flow rule, $\dot{\varepsilon}_c \propto \sigma^n$ with $n = 3.5$, implies that the creep response of a given structure is far more sensitive to the applied load than its elastic response. In this experiment we examine how this nonlinearity is reflected in the optimized topology itself. Returning to the purely structural cantilever of \Cref{sec:expts_validation}, with the traction applied at the center of the right edge, we sweep the applied traction magnitude across $\{100, 200, 400\}$ MPa while holding the service life (1 year), volume fraction ($V_* = 0.5$), and material properties fixed. Each case is independently optimized under the creep-aware objective.

\begin{figure}[h]
    \begin{center}
        \includegraphics[width=1.0\linewidth]{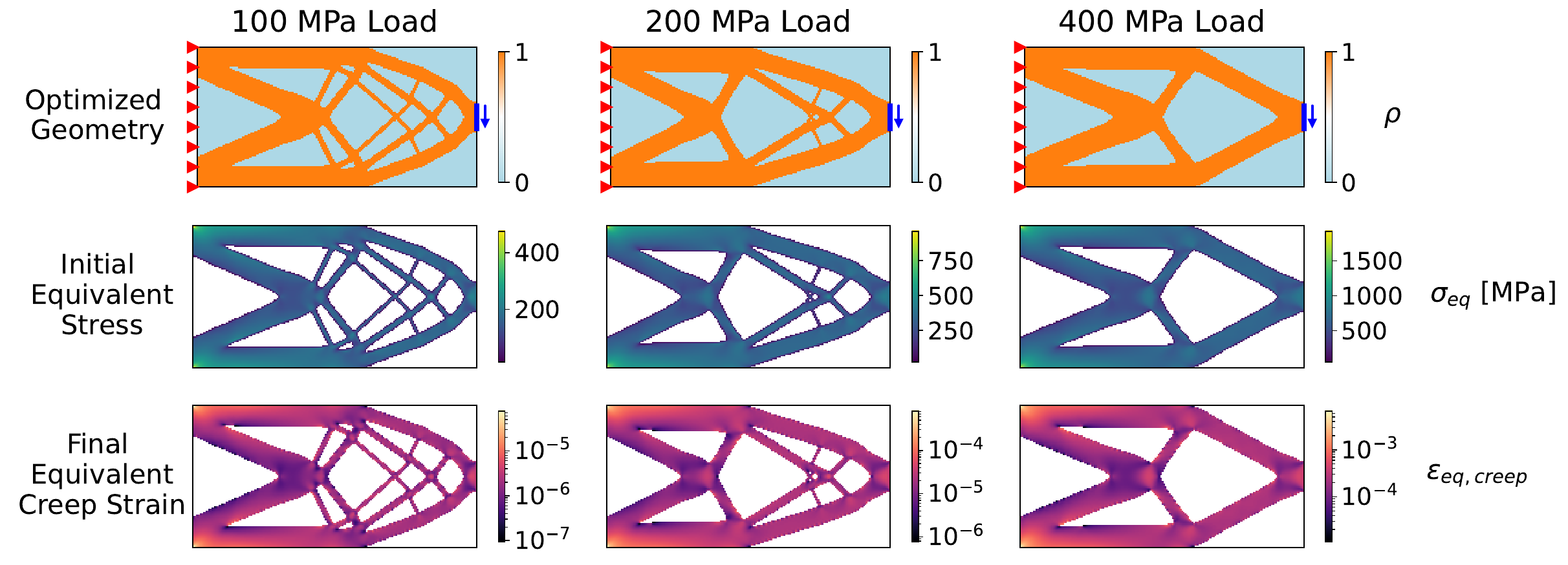}
        \caption{Effect of applied load magnitude on the optimized topology. \textit{Top}: Optimized material distributions for applied tractions of 100, 200, and 400 MPa. \textit{Middle}: Initial von Mises stress distribution. \textit{Bottom}: Final accumulated equivalent creep strain, plotted on a logarithmic scale. As the load increases, the optimizer consolidates material into fewer, thicker members, and the peak accumulated creep strain grows by roughly two orders of magnitude across the four-fold load range.}
        \label{fig:result_load_magnitude}
    \end{center}
\end{figure}

\Cref{fig:result_load_magnitude} shows the three optimized topologies alongside their initial von Mises stress and final accumulated creep strain distributions. Two trends stand out. First, as the applied load increases, the optimizer progressively consolidates material into fewer, thicker load-bearing members: the 100 MPa design features an intricate truss-like architecture with many slender members and internal substructure, the 200 MPa design retains the primary load paths but sheds much of the internal detailing, and the 400 MPa design converges to a compact form dominated by a few thick primary members. Second, while the peak initial stress scales roughly linearly with the applied traction, the peak accumulated creep strain grows by nearly two orders of magnitude across the four-fold load increase, from $\sim 10^{-5}$ at 100 MPa to $\sim 10^{-3}$ at 400 MPa, consistent with the $\sigma^n$ scaling of the Norton flow rate ($4^{3.5} \approx 128$). The optimized creep compliances for the three load cases are $4.29$, $91.0$, and $1897$ mJ, respectively.

This behavior is a direct consequence of the same stress-nonlinearity mechanism identified in the volume-fraction Pareto study of \Cref{sec:expts_pareto}. Because the creep rate scales as $\sigma^n$, the marginal creep penalty of any given stress concentration grows explosively with the load, and the optimizer responds by trading topological detail for reductions in peak sustained stress. At low loads, the stress in every member is well below the range where Norton flow becomes aggressive, and the optimizer can afford to distribute material across many fine members that primarily serve elastic stiffness. At high loads, even modest stress concentrations drive rapid creep accumulation, and the optimal strategy shifts toward increasing member cross-sections to suppress peak stresses at the expense of geometric intricacy.

This outcome is somewhat counterintuitive. Fine truss-like members introduce traction-free surfaces that serve as local stress relievers, redirecting load away from concentrations that would otherwise drive rapid Norton flow; a mechanism the optimizer readily exploits at moderate loads. The results here indicate that, once the applied load is large enough, this stress-relief benefit is outweighed by the need to reduce the peak sustained stress magnitude itself, and the optimizer accordingly abandons intricate substructure in favor of bulk cross-section. The cross-over between these two strategies is set by the nonlinearity of the flow rule: the stress-reduction benefit of added cross-section scales as $\sigma^{n}$, whereas the elastic benefit of stress-relief surfaces scales only linearly, so the former eventually dominates.

The broader insight is that, unlike elastic-compliance optimization where the optimal topology is largely load-invariant and the design can be scaled to accommodate different service loads, the creep-optimal topology depends qualitatively on the load magnitude. Components expected to operate under high sustained loads require a fundamentally different topological strategy than those designed for lower loads, and the applied load must accordingly be prespecified as a design input rather than treated as a post-hoc scaling parameter in creep-aware TO.

\subsection{Three-Dimensional Graded Alloy Turbine Blade}
\label{sec:expts_3D}

The preceding studies used the design variable field to redistribute a single solid material against void, treating the topology as the sole lever available to the optimizer. Real high-temperature components, however, are increasingly manufactured as functionally graded alloys in which the local chemistry, and therefore the local creep response, is itself a designable field \citep{reichardt_advances_2021}. In this final example we hold the geometry fixed and instead let the optimizer distribute two candidate alloys, Inconel~625 (IN625) and stainless steel~316 (SS316), across a three-dimensional turbine blade. Recent high-throughput experiments have demonstrated that creep laws can now be discovered across broad regions of alloy compositional space \citep{chen2026MLCreep}, making compositional design a realistic extension of the present framework to production materials rather than an abstract hypothetical.

We follow the linear radial-basis-function (RBF) interpolation scheme of \citet{Knapik2025} to construct spatially varying properties from a small set of base alloys. Each element carries a compositional design variable $\bm{\theta}_e = (\theta_e^{\text{IN625}}, \theta_e^{\text{SS316}})$, distinct from the scalar topology pseudodensity $\rho_e$ used previously, constrained to the simplex $\theta_e^{\text{IN625}} + \theta_e^{\text{SS316}} = 1$, so that the two entries represent complementary volume fractions of the base alloys. Every temperature-dependent property $P \in \{E, k\}$ and creep parameter $P \in \{A_0, n, Q\}$ is then evaluated as a linear RBF blend of the base-alloy values, with the RBF weights recovering pure-alloy properties at the compositional vertices and interpolating between them. Unlike the RAMP scheme of \Cref{eq:kappa_E_interpolation,eq:A_interpolation}, which penalizes intermediate densities to drive the design toward a binary field, the RBF blend is not penalized: intermediate compositions represent physical alloy mixtures rather than fictitious material states. We acknowledge that a linear blend is not a physically faithful model of alloy behavior at intermediate compositions, where phase formation, solute strengthening, and nonlinear diffusion-driven creep mechanisms are known to produce properties that deviate substantially from a simple rule-of-mixtures. We adopt the linear RBF blend here to keep the example self-contained and because dense compositional-property data across the IN625--SS316 space is not readily available; the framework itself is agnostic to the choice of interpolant, and the recent creep-law-discovery efforts of \citep{chen2026MLCreep} point toward the kind of composition-resolved datasets that would enable a physically calibrated replacement in future work. The base-alloy properties used in this study are summarized in \Cref{tab:blade_materials}; IN625 offers substantially greater creep resistance (its Norton coefficient $A_0$ is roughly eleven orders of magnitude smaller than SS316's), while SS316 is the more thermally conductive alloy.

\begin{table}[h]
    \centering
    \caption{Base-alloy properties used in the graded-blade study. Thermal conductivity $k$ and Young's modulus $E$ are temperature-dependent; representative cold- and hot-side values are shown. Norton creep parameters $(A_0, n, Q)$ are taken as temperature-independent per alloy, with the Arrhenius factor $\exp(-Q/RT)$ supplying the temperature dependence of the creep rate.}
    \label{tab:blade_materials}
    \begin{tabular}{lccccccc}
        \toprule
        & \multicolumn{2}{c}{$E$ [GPa]} & \multicolumn{2}{c}{$k$ [W/(mm$\cdot$K)]} & $A_0$ & $n$ & $Q$ \\
        \cmidrule(lr){2-3} \cmidrule(lr){4-5}
        Alloy & $250\,^\circ$C & $750\,^\circ$C & $250\,^\circ$C & $750\,^\circ$C & [MPa$^{-n}\!\cdot\!$s$^{-1}$] & [--] & [kJ/mol] \\
        \midrule
        IN625 & 195.4 & 161.5 & $1.32\times10^{-2}$ & $2.06\times10^{-2}$ & $10^{-34}$ & 4.5 & 275 \\
        SS316 & 185.7 & 141.0 & $1.66\times10^{-2}$ & $2.34\times10^{-2}$ & $10^{-23}$ & 5.5 & 230 \\
        \bottomrule
    \end{tabular}
\end{table}

\begin{figure}[h]
    \begin{center}
        \includegraphics[width=0.9\linewidth]{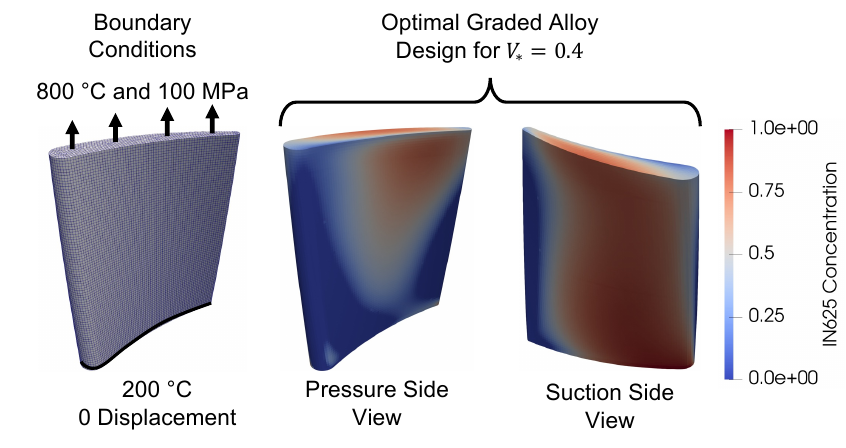}
        \caption{Three-dimensional graded-alloy turbine blade. \textit{Left}: blade geometry and boundary conditions, with the root clamped at $T_\text{cold} = 200\,^\circ$C, the tip held at $T_\text{hot} = 800\,^\circ$C, and a sustained $100~\text{MPa}$ traction applied to the top face. \textit{Right}: optimized IN625 concentration field for the graded design at $V_* = 0.4$, with the superalloy concentrated toward the hot tip and suction side.}
        \label{fig:result_blade_BC_design}
    \end{center}
\end{figure}

The blade geometry is discretized with a structured Hex8 mesh containing approximately 50k elements. The root (bottom face, $y = 0$) is fully clamped and held at $T_\text{cold} = 200\,^\circ$C, and the tip (top face, $y = 50~\text{mm}$) is held at $T_\text{hot} = 800\,^\circ$C, giving a strong through-span thermal gradient. A sustained traction of magnitude $100~\text{MPa}$ is applied to the top face, representing the aggregated centrifugal loading during steady operation. The service life is one year. The compositional field is initialized uniformly at $\bm{\theta}_e = (0.5, 0.5)$, and the optimizer minimizes the creep compliance subject to a volume-fraction constraint on the IN625 content, $\bar{\theta}^{\text{IN625}} \le V_*$, reflecting the practical incentive to conserve the more expensive superalloy. The blade geometry and boundary conditions are shown in \Cref{fig:result_blade_BC_design}, alongside the optimized IN625 concentration field for the graded design at $V_* = 0.4$.

\begin{figure}[h]
    \begin{center}
        \includegraphics[width=0.78\linewidth]{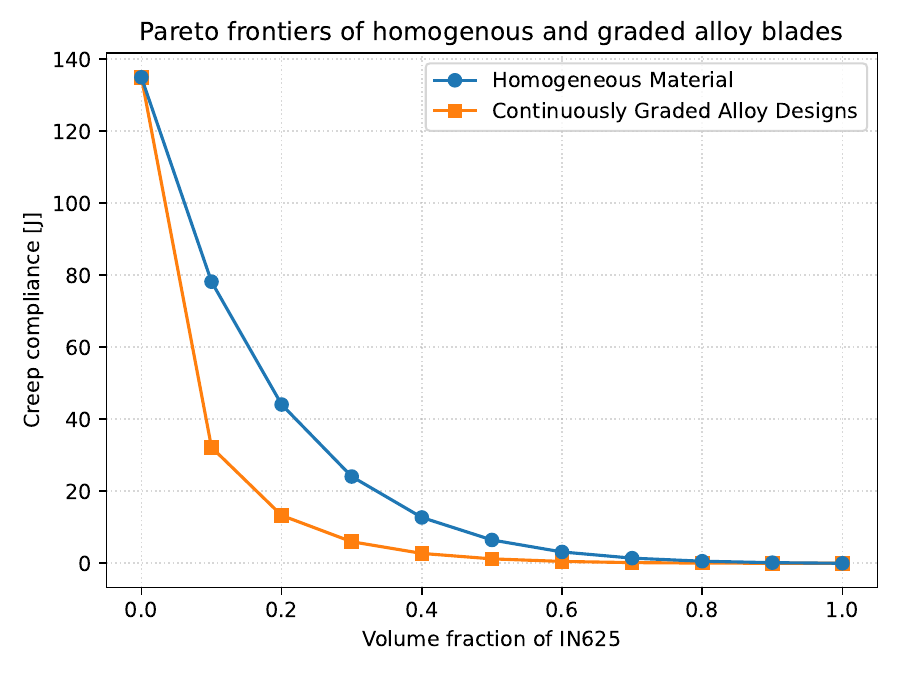}
        \caption{Pareto frontiers of creep compliance versus IN625 volume fraction for the 3D turbine blade. The blue curve corresponds to homogeneous blends in which the IN625/SS316 ratio is spatially uniform; the orange curve corresponds to optimizer-designed continuously graded compositions.}
        \label{fig:result_blade_pareto}
    \end{center}
\end{figure}

We sweep $V_*$ across $\{0.1, 0.2, \ldots, 0.9\}$ and compare the optimized graded compositions against a homogeneous baseline in which the IN625/SS316 ratio is spatially uniform at the same overall volume fraction. \Cref{fig:result_blade_pareto} shows the two Pareto frontiers on the plane of creep compliance versus IN625 volume fraction. Three observations stand out. First, the graded designs strictly Pareto-dominate the homogeneous blends across the full sweep, and the gap is largest in the lower range $V_* \in [0.1, 0.5]$ where a graded design at $V_* = 0.4$ attains a creep compliance ($\approx 2.5$ J) comparable to what the homogeneous baseline requires roughly $V_* = 0.6$ to achieve. In other words, roughly one third of the IN625 needed for a given creep-compliance target can be eliminated simply by allowing the alloy to be spatially graded. Second, both frontiers exhibit the same strongly convex knee already observed in the geometric Pareto study of \Cref{sec:expts_pareto}, reinforcing that the diminishing-returns behavior is a generic consequence of the Norton stress-nonlinearity and not an artifact of the design parameterization. Third, the endpoints of the two frontiers agree exactly, as they must: a pure-SS316 blade ($V_* = 0$) and a pure-IN625 blade ($V_* = 1$) are homogeneous by definition, so grading has no room to help. All of the value created by compositional design lives in the interior of the sweep.

The right panel of \Cref{fig:result_blade_BC_design} shows the optimized IN625 concentration for the graded design at $V_* = 0.4$. The superalloy is not distributed uniformly, nor even placed simply on the hot face; instead the optimizer concentrates IN625 along the hot tip and along the more highly stressed suction side, while the cooler root region and the lower-stressed pressure side are left almost entirely as SS316. This spatial pattern mirrors, in the compositional domain, the geometric shielding behavior identified in the thermal-gradient study of \Cref{sec:expts_thermal}: the optimizer places its creep-resistant resource preferentially where both the Arrhenius factor $\exp(-Q/RT)$ and the sustained stress are largest, because it is the product $A(T)\,\sigma^n$ that governs the local creep rate, and a reduction in the local creep coefficient buys the greatest reduction in accumulated creep strain precisely where this product is highest. The mechanism is identical to that of geometric material redistribution; only the physical meaning of the design variables have changed. Together with the results of \Cref{sec:expts_thermal}, this observation supports a general design principle for creep-limited components under thermal gradients: hot regions demand creep-resistant resource, whether supplied geometrically as additional cross-section or compositionally as superalloy.

Each of the 3D blade optimizations converged in roughly $30$ MMA iterations, requiring on the order of $25$ minutes of wall time on a node with an AMD EPYC~7413 CPU and an NVIDIA RTX~A6000 GPU. The bulk of this cost lies in the forward nonlinear solves of the time-stepping procedure (blue in \Cref{fig:method_flowchart}), which alternate between a sparse global equilibrium Newton-Raphson on the CPU and dense per-quadrature-point return-mapping Newton-Raphson (red) on the GPU; the adjoint sensitivity computation reduces to linear solves and is comparatively fast.

\section{Discussion and Conclusion}
\label{sec:discussion}

This work presents a gradient-based TO framework for minimizing thermo-structural creep deformation. The framework couples a steady-state thermal solve with a transient, nonlinear structural analysis governed by the Norton power-law secondary-creep model. Sensitivities are computed via reverse-mode automatic differentiation combined with the Implicit Function Theorem applied at each Newton-Raphson solve, yielding machine-precise gradients without manual adjoint derivations or backpropagation through iterative solver history.

The numerical experiments collectively establish four findings. First, the validation study (\Cref{sec:expts_validation}) demonstrates that optimizing for creep compliance yields measurably lower long-term permanent deformation than elastic-compliance designs, even when the two topologies appear visually similar at convergence. The divergence in transient displacement history confirms that the creep-aware objective captures time-dependent failure mechanisms that elastic objectives cannot. Second, the Pareto study (\Cref{sec:expts_pareto}) shows that the volume fraction constraint governs a strongly convex tradeoff between the material budget and the attainable creep compliance, with diminishing returns as the budget is relaxed. This convexity is a consequence of the $\sigma^n$ stress-nonlinearity of the Norton flow rule; the specific location of the knee observed in our experiments is problem-dependent and should not be taken as a general threshold. Third, the thermal coupling experiments (\Cref{sec:expts_thermal}) reveal that the Arrhenius temperature dependence of the creep coefficient is not merely a scalar amplification of the mechanical problem. When a spatial thermal gradient is present, the optimizer thickens members on the hot side to suppress the sustained stress there, exploiting the fact that a reduction in stress in the hot region is disproportionately valuable when $A(T)$ is large. Uniform temperature shifts, by contrast, only rescale the accumulated creep and do not drastically reshape the topology. Isothermal creep-aware designs will therefore systematically misplace material in thermally inhomogeneous environments. Finally, the load magnitude study (\Cref{sec:expts_loadmagnitude}) shows that, in sharp contrast to elastic-compliance optimization, the creep-optimal topology depends qualitatively on the applied load. As the load grows, the optimizer abandons intricate truss-like substructure in favor of fewer, thicker cross-sections that suppress peak sustained stress, because the marginal creep penalty of any stress concentration scales as $\sigma^n$ and eventually dominates the elastic stress-relief benefit of slender members. The design load must accordingly be treated as a specification rather than a scaling parameter.

The three-dimensional functionally-graded turbine blade in \Cref{sec:expts_3D} extends these observations into the compositional design domain, where the design variable is a continuous alloy concentration field rather than a density. Sweeping the IN625 volume fraction and comparing optimizer-designed graded compositions against spatially uniform blends, we find that graded designs strictly Pareto-dominate homogeneous ones across the full sweep. At $V_* = 0.4$ a graded design attains a creep compliance comparable to what the homogeneous baseline requires roughly $V_* = 0.6$ to achieve, eliminating roughly one third of the more expensive superalloy for a given creep-compliance target. The mechanism mirrors the geometric shielding of \Cref{sec:expts_thermal} in the compositional domain: the optimizer concentrates IN625 in the hottest regions where the Arrhenius factor is largest and any reduction in the local creep coefficient buys the greatest reduction in accumulated creep strain. The structural conclusions drawn from this example are conditional on the material model used: the linear RBF blend applied to the Norton parameters in graded regions is a hypothetical assumption because dense compositional creep data across the IN625-SS316 space is not yet available. As such datasets mature, validated material models can be substituted directly into the optimization pipeline.

The key contributions of this work are:
\begin{enumerate}
    \item \textbf{Thermo-mechanically coupled creep-aware TO:} To the best of our knowledge, this is the first TO framework to couple a thermal field with transient Norton creep optimization, capturing the Arrhenius temperature dependence of the creep coefficient. The experiments confirm that this coupling qualitatively reshapes the optimized topology in ways that isothermal creep models cannot capture.
    \item \textbf{Automatic adjoint-based sensitivity computation via differentiable simulation:} By implementing the full thermo-mechanical creep solver within JAX, the framework leverages reverse-mode automatic differentiation and the implicit function theorem (IFT) to compute exact design sensitivities. This eliminates the need for manual derivation and implementation of adjoint equations, reducing development effort while enabling extension to additional physics and constitutive models that depend on internal state variables.
    \item \textbf{Insights into thermo-structural creep design:} Our experiments reveal that creep-aware topologies differ qualitatively from elastic-compliance designs over long service horizons; that thermal gradients fundamentally reshape the optimal material distribution through the Arrhenius creep coefficient; that the applied load magnitude qualitatively alters the creep-optimal topology; and that these geometric insights carry over directly into the compositional domain, where the optimizer concentrates the more creep-resistant alloy in the hottest regions and most highly stressed regions.
\end{enumerate}

Several limitations of the present study point toward natural directions for future work. The structural model assumes small strains and linear elasticity; extending to an elasto-plastic constitutive model would capture rate-independent yielding, which can couple with creep at stress concentrations to accelerate deformation, while incorporating finite-strain kinematics would capture geometric nonlinearity relevant to the most demanding high-temperature applications. The Norton model captures secondary steady-state creep exclusively. Incorporating Norton-Bailey time hardening and Kachanov-Rabotnov continuum damage mechanics would enable modeling of primary creep transients and tertiary creep rupture, providing a more complete lifespan assessment tool. The experiments here use representative material parameters; calibrating the framework to experimentally validated creep databases for specific superalloys would enable direct engineering application. Finally, experimental validation of fabricated designs, tested under representative thermo-mechanical loading conditions, represents a critical next step toward deploying this framework in practice.

\backmatter

\section*{Declarations}
\begin{itemize}
  \item \textbf{Funding} This work was supported by the Defense Advanced Research Projects Agency (DARPA) Multiobjective Engineering and Testing of Alloy Structures (METALS) program project titled “RADICAL: Rapid Array DImple based Co-design of gradient materiaL and geometry” under cooperative agreement No. HR0011-24-2-0302. This work was also supported by the National Science Foundation grant 2219489 to WC and the Department of Defense Vannevar Bush Faculty Fellowship, N00014-19-1-2642, to JC.
  \item \textbf{Author Contributions} S.K. developed the conceptual framework, wrote the software, and conducted the experiments. S.K. and A.C. prepared the original manuscript. D.S. provided critical feedback and assisted in revising the manuscript. J.C., C.S., and W.C. supervised the project, acquired funding, and reviewed the final manuscript.
  \item \textbf{Conflict of Interest} The authors declare that they have no conflict of interest.
  \item \textbf{Replication of Results} The Python code is available at \href{https://github.com/ideal-nu/creep-topopt}{github.com/ideal-nu/creep-topopt}.
\end{itemize}

\bibliography{references}

\begin{appendices}
\crefalias{section}{appendix}
\section{Detailed Algorithms}\label{secA1}\label{sec:appendix_algorithms}

For completeness, we provide pseudocode for the four interconnected routines that comprise the framework summarized in \Cref{fig:method_flowchart}: the outer TO loop, the transient time-stepping analysis, the global Newton-Raphson equilibrium solver, and the local constitutive update at each Gauss point.


\Cref{alg:to_loop} outlines the outer TO loop. At each iteration, we evaluate the steady-state thermal problem and then the transient mechanical creep problem, compute the objective and constraint with their sensitivities via AD, and update the design using MMA.

\begin{algorithm}[H]
\caption{Topology Optimization (Design Loop)}
\label{alg:to_loop}
\begin{algorithmic}[1]
\Require Initial design $\bm{\rho}_0$, volume fraction $V_*$, max iterations $I_{\max}$, tolerance $\epsilon_{opt}$
\State Initialize $\bm{\rho} \leftarrow \bm{\rho}_0$, change $\leftarrow \infty$, iter $\leftarrow 0$
\While{change $> \epsilon_{opt}$ \textbf{and} iter $< I_{\max}$}
    \State iter $\leftarrow$ iter $+ 1$
    \State Filter densities: $\tilde{\bm{\rho}} \leftarrow \text{DensityFilter}(\bm{\rho})$ \Comment{\Cref{sec:method_designRepresentation}}
    \State Solve thermal problem: $\bm{K}_{th}(\tilde{\bm{\rho}})\,\bm{T} = \bm{F}_{th}$ \Comment{Solves \Cref{eq:thermal_PDE}}
    \State Solve structural problem: $(\bm{u}(t),\,\bm{\varepsilon}^{cr}(t)) \leftarrow \textsc{TimeSteppingLoop}(\tilde{\bm{\rho}}, \bm{T})$ \Comment{Call \Cref{alg:time_step}}
    \State Compute objective $J(\bm{u})$ \Comment{Evaluates \Cref{eq:objective_creep_compliance}}
    \State Compute constraint $g_v(\tilde{\bm{\rho}})$ \Comment{Evaluates \Cref{eq:volume_constraint_integral}}
    \State Compute sensitivities $\partial J/\partial\bm{\rho}$, $\partial g_v/\partial\bm{\rho}$ 
    \State Update $\bm{\rho}$ via MMA \Comment{Towards \Cref{eq:optimizationEquations}}
    \State change $\leftarrow \|\bm{\rho}_{new} - \bm{\rho}_{old}\|_\infty$
\EndWhile
\State \Return Optimized design $\bm{\rho}^*$
\end{algorithmic}
\end{algorithm}


\Cref{alg:time_step} details the transient structural analysis. It advances the state over $N_t$ time steps using an implicit backward Euler scheme, calling the global equilibrium solver at each increment.

\begin{algorithm}[H]
\caption{Transient Analysis (Time Stepping Loop)}
\label{alg:time_step}
\begin{algorithmic}[1]
\Require Filtered densities $\tilde{\bm{\rho}}$, nodal temperatures $\bm{T}$, time steps $\Delta t_1, \dots, \Delta t_{N_t}$
\State Initialize: $\bm{u}_0 \leftarrow \bm{0}$, $\bm{\varepsilon}^{cr}_0 \leftarrow \bm{0}$
\For{$s = 1$ to $N_t$}
    \State Predictor: $\bm{u}_s^{(0)} \leftarrow \bm{u}_{s-1}$
    \State $(\bm{u}_s,\,\bm{\varepsilon}^{cr}_s) \leftarrow \textsc{GlobalNewtonRaphson}(\bm{u}_s^{(0)},\,\bm{\varepsilon}^{cr}_{s-1},\,\bm{T},\,\Delta t_s,\,\tilde{\bm{\rho}})$ \Comment{Call \Cref{alg:global_nr}}
    \State Store: $\bm{u}(t_s) \leftarrow \bm{u}_s$, $\bm{\varepsilon}^{cr}(t_s) \leftarrow \bm{\varepsilon}^{cr}_s$
\EndFor
\State \Return Displacement and creep strain histories $\bm{u}(t)$, $\bm{\varepsilon}^{cr}(t)$
\end{algorithmic}
\end{algorithm}


\Cref{alg:global_nr} describes the global Newton-Raphson force equilibrium solver. At each time step it resolves the nonlinear structural equilibrium, calling the local constitutive update at each Gauss point.

\begin{algorithm}[H]
\caption{Global Newton-Raphson (Equilibrium Solver)}
\label{alg:global_nr}
\begin{algorithmic}[1]
\Require Guess $\bm{u}^{(0)}$, history $\bm{\varepsilon}^{cr}_{s-1}$, temps $\bm{T}$, $\Delta t$, densities $\tilde{\bm{\rho}}$, tolerance $\epsilon_{nr}$
\State Initialize: $l \leftarrow 0$, $R_{norm} \leftarrow \infty$, $\bm{u}^{(l)} \leftarrow \bm{u}^{(0)}$
\While{$R_{norm} > \epsilon_{nr}$}
    \State $\bm{K}_T \leftarrow \bm{0}$, $\bm{R} \leftarrow \bm{0}$
    \For{each element $e = 1$ to $N_e$}
        \For{each Gauss point $g$}
            \State Compute total strain: $\bm{\varepsilon}_{s} \leftarrow \bm{B}_g\,\bm{u}_e^{(l)}$
            \State $(\bm{\sigma}_s,\,\Delta\bm{\varepsilon}^{cr}) \leftarrow \textsc{LocalUpdate}(\bm{\varepsilon}_s,\,\bm{\varepsilon}^{cr}_{e,s-1},\,T_g,\,\Delta t,\,\tilde{\rho}_e)$ \Comment{Call \Cref{alg:local_update}}
        \EndFor
        \State Compute $\bm{f}_{int}^e \leftarrow \int_{\Omega_e} \bm{B}^T\bm{\sigma}_s\,d\Omega$ and $\bm{K}_T^e \leftarrow \partial\bm{f}_{int}^e/\partial\bm{u}_e$ \Comment{\Cref{eq:mechanical_internal_force}}
        \State Assemble: $\bm{R} \mathrel{+}= \bm{f}_{ext}^e - \bm{f}_{int}^e$, $\bm{K}_T \mathrel{+}= \bm{K}_T^e$
    \EndFor
    \State Apply boundary conditions to $\bm{R}$ and $\bm{K}_T$
    \State Solve: $\bm{K}_T\,\Delta\bm{u} = \bm{R}$ \Comment{\Cref{eq:mechanical_linearized_system}}
    \State Update: $\bm{u}^{(l+1)} \leftarrow \bm{u}^{(l)} + \Delta\bm{u}$; $R_{norm} \leftarrow \|\bm{R}\|$; $l \leftarrow l+1$
\EndWhile
\State \Return Converged $\bm{u}_s$, updated $\bm{\varepsilon}^{cr}_s$
\end{algorithmic}
\end{algorithm}


\Cref{alg:local_update} details the local constitutive update at each Gauss point. Given the current total strain, the previous creep strain, the local temperature, and the time increment, it returns the updated stress and creep strain by solving the implicit residual (\Cref{eq:creep_local_residual_tensor}) via Newton-Raphson with AD-computed Jacobians.

\begin{algorithm}[H]
\caption{Local Constitutive Update (Gauss Point)}
\label{alg:local_update}
\begin{algorithmic}[1]
\Require Total strain $\bm{\varepsilon}_s$, previous creep strain $\bm{\varepsilon}^{cr}_{s-1}$, temperature $T$, $\Delta t$, material/density $(\tilde{\rho}_e, A_0, Q, n, \alpha, E, \nu)$
\State Compute penalized properties: $E \leftarrow E_1\cdot\text{RAMP}_q(\tilde{\rho}_e)$, $A_0 \leftarrow A_{0,1}\cdot\text{RAMP}_q(\tilde{\rho}_e)^{-n}$ \Comment{\Cref{eq:kappa_E_interpolation,eq:A_interpolation}}
\State Compute creep coefficient: $A(T) \leftarrow A_0\exp(-Q/RT)$ \Comment{\Cref{eq:norton_creep_coeff}}
\State Compute thermal strain: $\bm{\varepsilon}^{th} \leftarrow \alpha(T - T_0)\bm{I}$
\State Define residual $\bm{\mathcal{R}}(\Delta\bm{\varepsilon}^{cr})$:
\Statex \quad $\bm{\varepsilon}^{el} \leftarrow \bm{\varepsilon}_s - \bm{\varepsilon}^{cr}_{s-1} - \Delta\bm{\varepsilon}^{cr} - \bm{\varepsilon}^{th}$
\Statex \quad $\bm{\sigma} \leftarrow \mathbb{E}_{E,\nu}:\bm{\varepsilon}^{el}$
\Statex \quad $\bm{\mathcal{R}} \leftarrow \Delta\bm{\varepsilon}^{cr} - \Delta t\cdot\dot{\bm{\varepsilon}}^{cr}(\bm{\sigma})$ \Comment{\Cref{eq:creep_local_residual_tensor,eq:norton_creep_strain_rate}}
\State Initialize: $\Delta\bm{\varepsilon}^{cr} \leftarrow \bm{0}$
\Repeat
    \State Compute Jacobian: $\bm{\mathcal{J}} \leftarrow \partial\bm{\mathcal{R}}/\partial\Delta\bm{\varepsilon}^{cr}$ via AD
    \State Update: $\Delta\bm{\varepsilon}^{cr} \leftarrow \Delta\bm{\varepsilon}^{cr} - \bm{\mathcal{J}}^{-1}\bm{\mathcal{R}}(\Delta\bm{\varepsilon}^{cr})$
\Until{$\|\bm{\mathcal{R}}\| < \epsilon_{local}$}
\State Update creep strain: $\bm{\varepsilon}^{cr}_s \leftarrow \bm{\varepsilon}^{cr}_{s-1} + \Delta\bm{\varepsilon}^{cr}$ \Comment{\Cref{eq:creep_strain_update}}
\State Recompute: $\bm{\sigma}_s \leftarrow \mathbb{E}_{E,\nu}:(\bm{\varepsilon}_s - \bm{\varepsilon}^{cr}_s - \bm{\varepsilon}^{th})$
\State \Return $\bm{\sigma}_s$, $\bm{\varepsilon}^{cr}_s$
\end{algorithmic}
\end{algorithm}

\end{appendices}

\end{document}